\documentclass[12pt]{article}
\usepackage{amssymb}
\usepackage{epsfig}
\usepackage{float}
\usepackage{graphicx}

\begin{document}
\begin{center}
{\bf M\"ossbauer Antineutrinos: Recoilless Resonant Emission and Absorption of Electron Antineutrinos}\footnote{Based on a lecture presented at the IVth International Pontecorvo Neutrino Physics School, September 26 - October 06, 2010, Alushta, Crimea, Ukraine.}
\end{center}
\begin{center}
Walter Potzel
\end{center}
\begin{center}
{\em Physik-Department E15, Technische Universit\"at M\"unchen,\\
D-85748 Garching, Germany}
\end{center}
\begin{abstract}
Basic questions concerning phononless resonant capture of monoenergetic electron antineutrinos (M\"ossbauer antineutrinos) emitted in bound-state $\beta$-decay in the $^{3}$H - $^{3}$He system are discussed. It is shown that lattice expansion and contraction after the transformation of the nucleus will drastically reduce the probability of phononless transitions and that various solid-state effects will cause large line broadening. As a possible alternative, the rare-earth system $^{163}$Ho - $^{163}$Dy is favoured. M\"ossbauer-antineutrino experiments could be used to gain new and deep insights into several basic problems in neutrino physics.

\end{abstract}

\section{Conventional M\"ossbauer spectroscopy}

Consider a nucleus in an excited state with energy $E$ and lifetime $\tau$. If this nucleus is free (not bound in a crystal) the $\gamma$ ray emitted upon deexcitation exhibits the energy $E-E_{R}$ where $E_{R}$ is the recoil energy of the nucleus after emitting the $\gamma$ ray. In addition, because of the thermal motion of the free nucleus the energy spread of the emitted $\gamma$ ray will usually be much larger than the natural linewidth $\Gamma=\hbar/\tau$ as determined by the time-energy uncertainty principle. In the reverse process, to excite such a free nucleus to the state $E$ a $\gamma$-ray energy of $E+E_{R}$ is necessary because the nucleus recoils after absorption of the $\gamma$ ray and, of course, the absorption spectrum will also be Doppler broadened by thermal motion. The recoil energy is given by $E_{R}=E^{2}/(2Mc^{2})$ where $M$ is the mass of the nucleus and $c$ is the speed of light in a vacuum. Doppler broadening due to thermal motion and the recoil energy $E_{R}$ are typically in the meV range.

It was a great achievement reached by R.L. M\"ossbauer \cite{Moessbauer} to discover that $E_{R}$ as well as Doppler broadening due to thermal motion can be avoided by imbedding the nuclei (atoms) in a crystalline lattice. An excited nucleus bound in a crystal can emit - with the probability $f$, called the Lamb-M\"ossbauer factor - a $\gamma$ ray with the full energy $E$ and the natural linewidth (see section 3.1).

To mention some examples of interesting M\"ossbauer transitions: $^{57}$Fe $(E=14.4$ keV, $\Gamma=4.3\cdot10^{-9}$ eV), $^{67}$Zn $(E=93.3$ keV, $\Gamma=4.8\cdot10^{-11}$ eV), $^{109}$Ag $(E=87.7$ keV, $\Gamma=1.2\cdot10^{-17} $ eV). The resonance cross section for a M\"ossbauer $\gamma$ ray which is emitted (with $E_{R}=0$) in a source and thereafter absorbed (again with $E_{R}=0$) by a target (absorber) is given by \cite{Moessbauer},\cite{Moessbauer1},\cite{Kells}

\begin{equation}\label{cross section}
\sigma_{R}^{\gamma}=2\pi\left(\frac{\lambda}{2\pi}\right)^{2} s^{2} f^{2} \frac{\Gamma}{\Gamma_{exp}} ~,
\end{equation}
where $\lambda$ is the wavelength of the $\gamma$ photon\footnote{Here we use the relation $\lambda=h/p$, where $h$ is Planck's constant and $p$ is the momentum.}; $s$ is a statistical factor which is determined by nuclear spins, isotopic abundance, etc. and can be considered to be of the order of unity. The experimental linewidth $\Gamma_{exp}$ takes line-broadening effects into account (see section 3.2). The maximal cross section $\sigma_{R}^{max}=2\pi\left(\frac{\lambda}{2\pi}\right)^{2}$ is only determined by $\lambda$ and is typically in the range between $10^{-17}$ and $10^{-19}$ cm$^2$ and thus very much larger than the cross section for weak interaction which is $\sim10^{-44}$ cm$^2$. The large cross section makes M\"ossbauer spectroscopy very interesting in many areas of physics. M\"ossbauer $\bar{\nu_{e}}$ - if they could be produced, e.g., by bound-state $\beta$-decay (see section 2) - would also exhibit these large resonance cross sections since M\"ossbauer $\bar{\nu_{e}}$ are characterized by low energies, where $\lambda$ is much is much larger than the dimensions of a nucleus. In this limit, the specific properties of the weak interaction come into play only via the natural linewidth $\Gamma$, i.e., the lifetime of the resonant state.

\section{Usual $\beta$-decay and bound-state $\beta$-decay}

In the usual continuum-state $\beta$-decay (C$\beta$), a neutron in a nucleus transforms into a proton and the electron (e$^-$) and electron-antineutrino ($\bar{\nu_{e}}$) are emitted into continuum states. This is a three-body process where the e$^-$ and the $\bar{\nu_{e}}$ exhibit broad energy spectra.

In the bound-state $\beta$-decay (B$\beta$), again a neutron in a nucleus transforms into a proton. The electron, however, is directly emitted into a bound-state atomic orbit \cite{Bahcall}. This is a two-body process. Thus the emitted $\bar{\nu_{e}}$ has a fixed energy $E_{\bar{\nu_{e}}}=Q+B_{z}-E_{R}$, determined by the Q value, the binding energy $B_{z}$ of the atomic orbit the electron is emitted into, and by the recoil energy $E_{R}$ of the atom formed after the decay.

The reverse process is also possible: an electron-antineutrino and an electron in an atomic orbit are absorbed by the nucleus and a proton is transformed into a neutron. Also this is a two-body process. The required energy of the antineutrino is given by $E'_{\bar{\nu_{e}}}=Q+B_{z}+E'_{R}$, where $E'_{R}$ is the recoil energy of the atom after the transformation of a proton into a neutron.

Due to the well-defined energies $E_{\bar{\nu_{e}}}$ and $E'_{\bar{\nu_{e}}}$, B$\beta$ has a resonant character \cite{Bahcall},\cite{Mika} which is partially destroyed by the recoil occurring after emission and absorption of the antineutrino. The resonance cross section (without M\"ossbauer effect) is given by \cite{Mika} $\sigma=4.18\cdot10^{-41}\cdot g_{0}^{2}\cdot \varrho(E_{\bar{\nu_{e}}}^{res})/ft_{1/2}$  cm$^{2}$, where $g_{0}=4\pi(\hbar/mc)^{3}\vert\psi\vert^{2}\approx4(Z/137)^{3}$ for low-Z, hydrogen-like wavefunctions $\psi$; \textit{m} is the electron mass, and $\varrho(E_{\bar{\nu_{e}}}^{res})$ is the resonant spectral density, i.e., the number of antineutrinos in an energy interval of 1MeV around $E_{\bar{\nu_{e}}}^{res}$. For a super-allowed transition, $ft_{1/2}\approx1000$.

The $^{3}$H - $^{3}$He system has been considered as a favourable example \cite{Kells}, \cite{RajuRag} for the observation of M\"ossbauer $\bar{\nu_{e}}$. The lifetime of $^{3}$H is $\tau=17.81$ y which corresponds to a natural linewidth $\Gamma=\hbar/\tau = 1.17\cdot 10^{-24}$ eV. The resonance energy is 18.59 keV; $ft_{1/2}\approx1000$. If gases of $^{3}$H and $^{3}$He are used at room temperature the profiles of the emission and absorption probabilities are Doppler broadened and both emission as well as absorption of the electron antineutrinos will occur with recoil. Thus the expected resonance cross-section (without M\"ossbauer effect) is still tiny, $\sigma\approx 1\cdot10^{-42}$ cm$^{2}$. However, recoilfree resonant antineutrino emission and absorption (M\"ossbauer effect) would increase the resonance cross-section by many orders of magnitude (see section 1). To prevent the recoil and avoid Doppler broadening, $^{3}$H as well as $^{3}$He have been considered to be imbedded in Nb metal lattices \cite{RajuRag}. It has to be taken into account that the ratio B$\beta/$C$\beta=\alpha=6.9\cdot10^{-3}$ with 80\% and 20\% of the B$\beta$ events proceeding via the atomic ground state and excited atomic states, respectively \cite{Bahcall}. In the case of the M\"ossbauer resonance only the fraction of transitions into (from) the ground-state atomic orbit is relevant, i.e., $\alpha\approx0.005$ in the source as well as in the absorber. Following eq. (\ref {cross section}), the resonance cross section for M\"ossbauer $\bar{\nu_{e}}$ can be written as \cite{Moessbauer},\cite{Kells}

\begin{equation}\label{cross section neutrinos}
\sigma_{R}^{\bar{\nu_{e}}}=2\pi\left(\frac{\lambda}{2\pi}\right)^{2} s^{2} \alpha^{2} f^{2} \frac{\Gamma}{\Gamma_{exp}} \approx1.8\cdot10^{-22}s^{2}f^{2}  \frac{\Gamma}{\Gamma_{exp}}~~ cm^{2}
\end{equation}

i.e., for the $^{3}$H - $^{3}$He system, $\sigma_{R}^{\bar{\nu_{e}}}\approx1.8\cdot10^{-22}$ cm$^{2}$, if $s^{2}\approx1$, $f^{2}\approx1$, and $\Gamma_{exp}=\Gamma = 1.17\cdot10^{-24}$ eV (no line broadening).

In the following, the $^{3}$H - $^{3}$He system in Nb metal will be discussed as an example in more detail.

\section{M\"ossbauer $\bar{\nu_{e}}$: Basic questions and answers}
\subsection{Phononless transitions}
To make the M\"ossbauer effect possible lattice excitations have to be avoided. With $\bar{\nu_{e}}$ two kinds of lattice-excitation processes have to be considered:
\begin{enumerate}
\item \textit{Momentum transfer} due to emission/capture of a $\bar{\nu_{e}}$. This process leads to the recoilfree fraction $f_{r}$, well known from conventional M\"ossbauer spectroscopy. In the Debye model and in the limit of very low temperatures $T$, the recoilfree fraction $f_r$ is given by

\begin{equation}\label{recoilfree fraction}
f_{r}(T\rightarrow0)=exp \left\{-\frac{E^{2}}{2Mc^{2}}\cdot\frac{3}{2k_{B}\theta}\right\},
\end{equation}
where $\theta$ is the effective Debye temperature, $k_{B}$ is the Boltzmann constant, and $\frac{E^{2}}{2Mc^{2}}$ is the recoil energy which would be transmitted to a free atom of mass $M$. With $\theta\approx800K$ for the $^{3}$H - $^{3}$He system \cite{WalterPotzel1}, $f_{r}^2\approx0.07$ for $T\rightarrow0$.
\item \textit{Lattice expansion and contraction.} This process which is not present with conventional M\"ossbauer spectroscopy, occurs when the nuclear transformation takes place during which the $\bar{\nu_{e}}$ is emitted or absorbed: The $\bar{\nu_{e}}$ itself takes part in the nuclear processes transforming one chemical element into a different one. $^{3}$H and $^{3}$He are differently bound in the Nb lattice and use different amounts of lattice space. Thus, the lattice will expand or contract when the weak processes occur. The lattice deformation energies for $^{3}$H and $^{3}$He in the Nb lattice are $E_L^{H}=0.099$ eV and $E_L^{He}=0.551$ eV, respectively \cite{Puska}. Assuming again an effective Debye temperature of $\theta\approx800$ K one can estimate - in analogy to the situation with momentum transfer - the probability $f_L$ that this lattice deformation will {\em not} cause lattice excitations:

\begin{equation}\label{lattice deformation}
f_L\approx exp \left\{-\frac{E_L^{^{3}He}-E_L^{^{3}H}}{k_{B}\theta}\right\}\approx1\cdot10^{-3}.
\end{equation}
Thus, the total probability for phononless emission and consecutive pho\-non\-less capture of $\bar{\nu_{e}}$ is
\begin{equation}\label{phononless probability}
f^2=f_r^2\cdot f_L^2\approx7\cdot10^{-8},
\end{equation}
which is tiny and makes a real experiment with the $^{3}$H - $^{3}$He system extremely difficult.

\end{enumerate}

\subsection{Linewidth}
With M\"ossbauer $\bar{\nu_{e}}$, three types of line-broadening effects \cite{Balko},\cite{Coussement},\cite{WalterPotzeletal} are important.
\begin{enumerate}
\item \textit{Homogeneous broadening} is caused by electromagnetic relaxation, e.g., by spin-spin interactions between the nuclear spins of $^{3}$H and $^{3}$He and with the spins of the Nb nuclei. Contrary to the claim in Ref. \cite{RaghavanPRL}, such magnetic relaxations are stochastic processes. They lead to sudden, irregular transitions between hyperfine-split states originating, e.g., from magnetic moments (spins) of many neighbouring nuclei. Thus the wave function of the emitted particle (photon, $\bar{\nu_{e}}$) is determined by random (in time) frequency changes which cause line broadening characterized by the time-energy uncertainty relation \cite{WalterPotzel1}. As a consequence, the broadened line \textit{cannot} be decomposed into multiple sharp lines. For the system $^{3}$H - $^{3}$He in Nb metal, $\Gamma_{exp}\approx4\cdot10^{13}\Gamma$, thus $\left(\frac{\Gamma}{\Gamma_{exp}}\right)_{h}\approx2.5\cdot10^{-14}$ \cite{WalterPotzel1},\cite{WalterPotzelPRL},\cite{PotzelWagner}.

\item \textit{Inhomogeneous broadening} is caused by stationary effects in an imperfect lattice, e.g., by impurities and lattice defects, and, in particular, by the random distributions of $^{3}$H and $^{3}$He on the tetrahedral interstitial sites in Nb metal. Thus the periodicity of the lattice is destroyed and the binding energies $E_B$ of $^{3}$H and $^{3}$He in the Nb lattice will vary. Typical values for $E_B$ are in the eV range \cite{Puska}. In conventional M\"ossbauer spectroscopy with photons, using the best single crystals, such variations in $E_B$ cause energy shifts of $\sim10^{-12}$ eV. Since, in the nuclear transformations, the $\bar{\nu_{e}}$ energy is \textit{directly} affected by $E_B$, one has to expect that the variations of the $\bar{\nu_{e}}$ energy are much larger than $10^{-12}$ eV, probably in the $10^{-6}$ eV regime. Thus, inhomogeneous line broadening is estimated to give $\left(\frac{\Gamma}{\Gamma_{exp}}\right)_{i}\ll10^{-12}$ \cite{WalterPotzel1},\cite{WalterPotzelPRL}, probably $\left(\frac{\Gamma}{\Gamma_{exp}}\right)_{i}\approx10^{-18}$ \cite{PotzelWagner}.

\item Another contribution to line broadening in an imperfect lattice is caused by \textit{relativistic effects}. An atom vibrating around its equilibrium position in a lattice exhibits a mean-square velocity $\left\langle v^{2}\right\rangle$. According to Special Relativity Theory this results in a time-dilatation, $t'=t/\sqrt{1-\left\langle v^{2}\right\rangle/c^{2}}$, i.e., a reduction in frequency (energy $E$) of the $\bar{\nu_{e}}$: $E'=E\sqrt{1-\left\langle v^{2}\right\rangle/c^{2}}\approx E(1-\left\langle v^{2}\right\rangle/(2c^{2}))$. Thus, $\Delta E\approx-\left\langle v^{2}\right\rangle E/(2c^{2})$. Being proportional to $\left\langle v^{2}\right\rangle/c^{2}$, this reduction is called second-order Doppler shift (SOD). Due to the zero-point motion (zero-point energy) the contribution is present even in the low-temperature limit for $T\rightarrow0$. The essential point is, that a variation of the binding energies $E_B$ of $^{3}$H and $^{3}$He in an imperfect lattice will result in a variation of the effective Debye temperatures and thus also in a variation of the zero-point energies. If the effective Debye temperature varies by only 1 K, $(\Delta E/E)\approx2\times10^{-14}$, corresponding to a lineshift of $3\times10^{14}$ times the natural width $\Gamma$. Thus, this broadening effect is expected to give $\left(\frac{\Gamma}{\Gamma_{exp}}\right)_{SOD}\approx3\cdot10^{-15}$ comparable to or even smaller than $\left(\frac{\Gamma}{\Gamma_{exp}}\right)_{h}$ \cite{WalterPotzel1}. The broadening effect $\left(\frac{\Gamma}{\Gamma_{exp}}\right)_{SOD}$ is important but less serious than $\left(\frac{\Gamma}{\Gamma_{exp}}\right)_{i}$, the latter being mainly due to variations of $E_B$ as discussed above.

\end{enumerate}

Thus, there exist very serious difficulties to observe M\"ossbauer $\bar{\nu_{e}}$ with the system $^{3}$H - $^{3}$He in Nb metal: The probability of phonon-less emission and detection might be very low, $f^2\approx7\cdot10^{-8}$; homogeneous and, in particular, inhomogeneous line broadening may cause $\left(\frac{\Gamma}{\Gamma_{exp}}\right)_{i}$ to be as small as $\left(\frac{\Gamma}{\Gamma_{exp}}\right)_{i}\approx10^{-18}$. This is mainly due to the fact that inhomogeneities in an imperfect lattice \textit{directly} influence the energy of the $\bar{\nu_{e}}$ \cite{PotzelWagner}. In addition to the basic principles we have just discussed also enormous technological difficulties can be expected \cite{Schiffer} and possible alternatives to the $^{3}$H - $^{3}$He system should be searched for \cite{PotzelWagner}.

\section{Alternative system: $^{163}$Ho - $^{163}$Dy}

The system $^{163}$Ho - $^{163}$Dy has been considered as the next-best case \cite{Kells},\cite{PotzelWagner}. Compared to $^{3}$H - $^{3}$He in Nb metal, there exist three major advantages:
\begin{enumerate}
\item The Q value of 2.6 keV (i.e., the $\bar{\nu_{e}}$ energy) is very low, the mass of the rare earth nuclei is large. As a consequence, according to eq. (\ref{recoilfree fraction}), the recoilfree fraction is expected to be $f_{r}\approx1$.

\item The rare earth series is characterized by different numbers of electrons in the 4f shell, the outer 6s, 6p shells remaining to a large extent unchanged. Therefore, the chemical behaviour due to the bonding electrons of outer shells is altered only very slightly. Thus also the deformation energies of $^{163}$Ho and $^{163}$Dy can be expected to be similar. For this reason, the probability $f$ of phononless transitions could be larger than for the $^{3}$H - $^{3}$He system by $\sim7$ orders of magnitude.

\item Due to the similar chemical behaviour, rare-earth systems can also be expected to be less sensitive to \textit{variations of the binding energies.} In addition, because of the large mass, the relativistic effects mentioned in section 3.2 will be smaller.
\end{enumerate}

The main disadvantage will be the large magnetic moments due to the 4f electrons of the rare-earth atoms. Electronic magnetic moments are typically three orders of magnitude larger than nuclear moments. Thus, line broadening effects due to magnetic relaxation phenomena will be decisive for a successful observation of M\"ossbauer $\bar{\nu_{e}}$. However, conventional M\"ossbauer spectroscopy gathered a lot of information on the behaviour of rare-earth systems in the past. In particular, it has been shown that an experimental linewidth of $\Gamma_{exp}\approx5\cdot10^{-8}$ eV for the 25.65 keV M\"ossbauer resonance in $^{161}$Dy can be reached (see section 5). Thus it is possible to test and improve promising source and target materials by investigations using rare-earth M\"ossbauer $\gamma$ transitions and - in this way - compensate the lack of technological knowledge concerning the fabrication of high-purity materials, preferably single crystals, in kg quantities \cite{Kells}. Certainly, further feasibility studies have to be performed.

\section{Interesting experiments}

If the emission and absorption of M\"ossbauer $\bar{\nu_{e}}$ could be observed successfully, several basic questions and interesting experiments could be addressed \cite{WalterPotzel1}:

\begin{enumerate}
\item The states of the neutrinos $\nu_{e}, \nu_{\mu}$, and $\nu_{\tau}$ are characterized by a coherent superposition of mass eigenstates and can change their identity when they propagate in space and time, a process called flavour oscillation \cite{Kayser},\cite{Beuthe},\cite{Bilenkyetal1},\cite{Akhmedov},\cite{Bilenkyetal2},\cite{Kopp}. M\"ossbauer $\bar{\nu_{e}}$ are characterized by a very narrow energy distribution. The question has been asked: Do M\"ossbauer $\bar{\nu_{e}}$ oscillate? Considering the evolution of the neutrino state in \textit{time only}, according to the Schr\"odinger equation neutrino oscillations are characterized as a non-stationary phenomenon, and M\"ossbauer $\bar{\nu_{e}}$ would not oscillate because of their extremely narrow energy distribution: As stated by the time-energy uncertainty relation $\Delta t\cdot\Delta E\geq\hbar$, the time interval $\Delta t$ required for a system (consisting, e.g., of a superposition of mass eigenstates) to change significantly is given by $\Delta t\geq\hbar/\Delta E$. If the uncertainty in energy $\Delta E$ is small, as in the case of M\"ossbauer $\bar{\nu_{e}}$, $\Delta t$ will be large. An evolution of the neutrino state in \textit{space and time}, however, would make oscillations possible in both the non-stationary and also in the stationary (M\"ossbauer $\bar{\nu_{e}}$) case. In this way, M\"ossbauer $\bar{\nu_{e}}$ experiments could lead to a better understanding of the true nature of neutrino oscillations \cite{Bilenkyetal1},\cite{Akhmedov},\cite{Bilenkyetal2}. We now discuss this aspect in greater detail.

Considering the evolution of the $\bar{\nu_{e}}$ state in time only, M\"ossbauer $\bar{\nu_{e}}$ oscillations with an oscillation length determined by $\Delta m^{2}_{23}$ will \textit{not} be observed if the relative energy uncertainty fulfills the relation \cite{Bilenkyetal2}

\begin{equation}\label{energy uncertainty}
\frac{\Delta E}{E}\ll\frac{1}{4}\frac{\Delta m_{23}^{2}c^{4}}{E^{2}}
\end{equation}

where $\Delta m^{2}_{23}=m^{2}_{3}-m^{2}_{2}\approx2.4\cdot10^{-3}$ eV$^{2}$ is the 'large' atmospheric mass-squared difference.

For the $^{3}$H - $^{3}$He system ($E=18.6$ keV) this would mean $\left(\frac{\Delta E}{E}\right)_{H-He}\ll 1.8\cdot10^{-12}$ which - according to section 3.2 - most probably cannot be reached due to inhomogeneous line broadening. However, for the $^{163}$Ho - $^{163}$Dy system, eq. (\ref{energy uncertainty}) requires $\left(\frac{\Delta E}{E}\right)_{Ho-Dy}\ll 9.2\cdot10^{-11}$ or $\Delta E\ll 2.4\cdot10^{-7}$ eV.

For the 25.65 keV $\gamma$-transition in $^{161}$Dy (lifetime $\tau=29.1\cdot 10^{-9}$ s, i.e., natural width $\Gamma=\hbar/\tau\approx2.26\cdot 10^{-8}$ eV) an experimental linewidth of $\Gamma_{exp}\approx5\cdot10^{-8}$ eV has been achieved \cite{Greenwood}, which is $\sim5$ times below the limit $\Delta E\lesssim 2.4\cdot10^{-7}$ eV. Thus, the $^{163}$Ho - $^{163}$Dy system will probably make it possible to perform M\"ossbauer $\bar{\nu_{e}}$ experiments, and it looks incouraging that the question if M\"ossbauer $\bar{\nu_{e}}$ oscillate can be answered experimentally. For $\Gamma_{exp}\approx5\cdot10^{-8}$ eV, a significant change of the $\bar{\nu_{e}}$ state in time can occur only very slowly leading to a long oscillation path $L_{change}$ since the $\bar{\nu_{e}}$ is ultrarelativistic:

\begin{equation}\label{changeduringtime}
L_{change}\simeq c\cdot\frac{\hbar}{\Gamma_{exp}}\cdot 2\pi.
\end{equation}

Thus, for the $^{163}$Ho - $^{163}$Dy system, $L_{change}\approx25$ m.

For an evolution of the $\bar{\nu_{e}}$ state in space and time, however, the oscillation length is given by

\begin{equation}\label{oscilength}
L_{0}=2.48\frac{E}{|\Delta m^{2}|}
\end{equation}

where $L_{0}$ is given in meters, when the neutrino energy $E$ and the mass-squared difference $|\Delta m^{2}|$ are in units of MeV and eV$^{2}$, respectively. With $E=2.6$ keV for the $^{163}$Ho - $^{163}$Dy system, and $\Delta m^{2}_{23}\approx2.4\cdot10^{-3}$ eV$^{2}$, we obtain $L_{0}\approx2.6$ m, considerably shorter than $L_{change}$. Thus in such a M\"ossbauer-neutrino experiment with $\Gamma_{exp}\approx5\cdot10^{-8}$ eV and if the evolution occurs in time only, instead of $L_{0}$ the much longer $L_{change}$ would be observed.

\item If M\"ossbauer-neutrino oscillations do occur, either due to the evolution of the neutrino state in \textit{space and time} or if $\Gamma_{exp}$ is so large that eq. (\ref{energy uncertainty}) is no longer fulfilled, ultra-short base lines (see eq. (\ref{oscilength})) could be used to determine oscillation parameters. For example, for the determination of the still unknown mixing angle $\Theta_{13}$, a base line of only $\sim1$ m (instead of $\sim$1500 m as required for reactor neutrinos \cite{2chooz}) would be sufficient. In addition, very small uncertainties for $\Theta_{13}$ and accurate measurements of $\Delta m^{2}_{12}=m^{2}_{2}-m^{2}_{1}$ and $\Delta m^{2}_{23}=m^{2}_{3}-m^{2}_{2}$ could become possible \cite{Minakata}.

\item At present, only upper limits for the neutrino masses are known. In addition, the mass spectrum (hierarchy), i.e., the question if the small mass splitting belongs to the two lightest or to the two heaviest neutrinos, could not yet be determined. M\"ossbauer $\bar{\nu_{e}}$ of the $^{163}$Ho - $^{163}$Dy system could settle the question concerning the mass spectrum using a base line of $\sim$ 40 m. As shown in Refs. \cite{Minakata1},\cite{Parke}, in such an experiment, the superposition of two oscillations with different frequencies can be observed: a low-frequency (so-called solar neutrino) oscillation driven by $\Delta m^{2}_{12}=m^{2}_{2}-m^{2}_{1}\approx7.6\cdot10^{-5}$ eV$^{2}$ and a high-frequency (so-called atmospheric neutrino) oscillation driven by $\Delta m^{2}_{23}=m^{2}_{3}-m^{2}_{2}\approx2.4\cdot10^{-3}$ eV$^{2}$. In the case of the normal mass spectrum (the small mass splitting belongs to the two lightest neutrinos) the phase of the atmospheric-neutrino oscillation advances, whereas it is retarded for the inverted mass spectrum (the small mass splitting belongs to the two heaviest neutrinos), by a measurable amount for every solar-neutrino oscillation.

\item Oscillating M\"ossbauer $\bar{\nu_{e}}$ could be used to search for the conversion $\bar{\nu_{e}}$ $\rightarrow\nu_{sterile}$ \cite{Kopeikin} involving additional mass eigenstates. Sterile neutrinos ($\nu_{sterile}$) would not show the weak interaction of the standard model of elementary particle interactions. Therefore such a conversion would manifest itself by the disappearance of $\bar{\nu_{e}}$. The results of the LSND (Liquid Scintillator Neutrino Detector) experiment \cite{LSND} indicate a mass splitting of $\Delta m^{2}\approx1$ eV$^2$ \cite{RajuRag}. In this context, several experiments have been performed by the MiniBooNE collaboration. However, the results are not conclusive \cite{MiniBooNE}. For a value of $\Delta m^{2}\approx1$ eV$^2$, the oscillation length for M\"ossbauer $\bar{\nu_{e}}$ of 2.6 keV would be only $\sim1$ cm!

\end{enumerate}

\section{Conclusions}

Several basic aspects of the $^{3}$H - $^{3}$He system have been considered for a possible observation of M\"ossbauer $\bar{\nu_{e}}$. It will not be possible to reach the natural linewidth because of homogeneous broadening - due to stochastic relaxation processes - and inhomogenous broadening mainly due to the variations of binding energies and their \textit{direct} influence on the energy of the $\bar{\nu_{e}}$. In addition, the probability for phononless emission and detection will drastically be reduced due to lattice expansion and contraction after the transformation of the nucleus. The observation of M\"ossbauer $\bar{\nu_{e}}$ of the system $^{3}$H - $^{3}$He imbedded in Nb metal, will most probably \textit{not} be possible. The rare-earth system $^{163}$Ho - $^{163}$Dy offers several advantages, in particular a large probability of phononless emission and detection. However, magnetic relaxation processes and technological requirements still have to be investigated in detail to decide if the $^{163}$Ho - $^{163}$Dy system is a successful alternative. In conventional M\"ossbauer experiments with the 25.65 keV $\gamma$ transition in $^{161}$Dy an experimental linewidth of $\Gamma_{exp}\approx5\cdot10^{-8}$ eV was achieved. If such a value could also be reached with M\"ossbauer $\bar{\nu_{e}}$ of the $^{163}$Ho - $^{163}$Dy system, highly interesting experiments concerning basic questions in neutrino physics could be performed.

\textbf{Acknowledgments}\\
It is a great pleasure to thank S. Bilenky, Joint Institute for Nuclear Research, Dubna, Russia, for numerous fruitful discussions. This work was supported by funds of the Deutsche For\-schungsgemeinschaft DFG (Transregio 27: Neutrinos and Beyond), the Munich Cluster of Excellence (Origin and Structure of the Universe), and the  Maier-Leibnitz-Laboratorium (Garching).


\begin{thebibliography}{99}
\bibitem{Moessbauer} R.L. M\"ossbauer, Z. Physik \textbf{151}, 124 (1958).
\bibitem{Moessbauer1} R.L. M\"ossbauer, Hyperfine Interact. \textbf{126}, 1 (2000).
\bibitem{Kells} W.P. Kells and J.P. Schiffer, Phys. Rev. C\textbf{28}, 2162 (1983).
\bibitem{Bahcall} J.N. Bahcall, Phys. Rev. \textbf{124}, 495 (1961).
\bibitem{Mika} L.A. Mika\'elyan, B.G. Tsinoev, and A.A. Borovoi, Sov. J. Nucl. Phys.  \textbf{6}, 254 (1968).
\bibitem{RajuRag} R.S. Raghavan, hep-ph/0601079 v3; arXiv: 0805.4155 [hep-ph] and 0806.0839 [hep-ph].

\bibitem{WalterPotzel1} W. Potzel, J. Phys.: Conf. Ser. \textbf{136}, 022010 (2008), arXiv: 0810.2170 [hep-ph].
\bibitem{Puska} M.J. Puska and R.M. Nieminen, Phys. Rev. B\textbf{10}, 5382 (1984).
\bibitem{Balko} B. Balko, I.W. Kay, J. Nicoll, and J.D. Silk, Hyperfine Interact.  \textbf{107}, 283 (1997).
\bibitem{Coussement} R. Coussement, G. S'heeren, M. Van Den Bergh, and P. Boolchand, Phys. Rev. B\textbf{45}, 9755 (1992).
\bibitem{WalterPotzeletal} W. Potzel et al., Hyperfine Interact. \textbf{72}, 197 (1992).
\bibitem{RaghavanPRL} R.S. Raghavan, Phys. Rev. Lett. \textbf{102}, 091804 (2009).
\bibitem{WalterPotzelPRL} W. Potzel and F.E. Wagner, Phys. Rev. Lett. \textbf{103}, 099101 (2009), arXiv: 0908.3985 [hep-ph].
\bibitem{PotzelWagner} W. Potzel and F.E. Wagner, in \textit{The Rudolf M\"ossbauer Story}, Springer-Verlag Berlin, to be published in 2011.
\bibitem{Schiffer} J.P. Schiffer, Phys. Rev. Lett. \textbf{103}, 099102 (2009).

\bibitem{Kayser} B. Kayser, Phys. Rev. D\textbf{24}, 110 (1981).
\bibitem{Beuthe} M. Beuthe, Phys. Rep. \textbf{375}, 105 (2003) and references therein.
\bibitem{Bilenkyetal1} S.M. Bilenky, F. von Feilitzsch, and W. Potzel, J. Phys. G: Nucl. Part. Phys. \textbf{34}, 987 (2007); J. Phys. G: Nucl. Part. Phys. \textbf{36}, 078002 (2009), arXiv: 0804.3409 [hep-ph].
\bibitem{Akhmedov} E.Kh. Akhmedov, J. Kopp, and M. Lindner, J. High Energy Phys. \textbf{0805}, 005 (2008), arXiv: 0802.2513 [hep-ph]; J. Phys. G: Nucl. Part. Phys. \textbf{36}, 078001 (2009), arXiv: 0803.1424v2 [hep-ph].

\bibitem{Bilenkyetal2} S.M. Bilenky, F. von Feilitzsch, and W. Potzel, J. Phys. G: Nucl. Part. Phys. \textbf{35}, 095003 (2008), arXiv: 0803.0527 v2 [hep-ph].
\bibitem{Kopp} J. Kopp, J. High Energy Phys. \textbf{0906}, 049 (2009), arXiv: 0904.4346 [hep-ph].

\bibitem{Greenwood} N.N. Greenwood and T.C. Gibb, \textit{M\"ossbauer Spectroscopy,} Chapman and Hall Ltd, London 1971.

\bibitem{2chooz}DOUBLE CHOOZ collaboration,
F. Ardellier et al., arXiv: hep-ex/0606025; T. Lasserre {\it
Europhys. News} \textbf{38} N4, 20 (2007); T. Kawasaki et al.,
{\it AIP Conf. Proc.} \textbf{981}, 202 (2008).
\bibitem{Minakata} H. Minakata and S. Uchinami, New J. Phys. \textbf{8}, 143 (2006), hep-ph/0602046.
\bibitem{Minakata1} H. Minakata, H. Nunokawa, S.J. Parke, and R. Zukanovich Funchal, Phys. Rev. D\textbf{76}, 053004 (2007), arXiv: hep-ph/0701151.
\bibitem{Parke} S.J. Parke, H. Minakata, H. Nunokawa, and R. Zukanovich Funchal, Nucl. Phys. Proc. Suppl. \textbf{188}, 115 (2008), arXiv: 0812.1879 [hep-ph].
\bibitem{Kopeikin} V. Kopeikin, L. Mikaelyan, and V. Sinev, hep-ph/0310246v2.
\bibitem{LSND} C. Athanassopoulos et al., LSND collaboration, Phys. Rev. Lett. \textbf{81}, 1774 (1998).
\bibitem{MiniBooNE} MiniBooNE collaboration, A.A. Aguilar-Arevalo et al., Phys. Rev. Lett. \textbf{98}, 231801 (2007); Phys. Rev. Lett. \textbf{102}, 101802 (2009), arXiv: 0812.2243 and 0901.1648; Phys. Rev. Lett. \textbf{105}, 181801 (2010), arXiv: 1007.1150v3 [hep-ex].






\end{thebibliography}
\end{document}